\documentclass{epl}
\usepackage{amsmath}
\title{Nonequilibrium phase transition in surface growth} 
\author{Buddhapriya CHAKRABARTI\inst{1,3}\thanks{E-mail: \email{buddho@physics.iisc.ernet.in}} \and Chandan DASGUPTA\inst{2,1,3}\thanks{E-mail: \email{cdgupta@physics.umd.edu}}}
\shortauthor{B. Chakrabarti \etal}
\institute{
\inst{1} Centre for Condensed Matter Theory, Department of Physics,
Indian Institute of Science, Bangalore 560012, India.\\
\inst{2} Department of Physics, Condensed Matter Theory Center, University of
Maryland, College Park, MD 20742-4111.\\
\inst{3} Condensed Matter Theory Unit, JNCASR, Bangalore 560064, India.} 
\pacs{81.10.Aj}{Theory and models of crystal growth; physics of crystal 
growth, crystal morphology and orientation.}
\pacs{81.15.Hi}{Molecular, atomic, ion, and chemical beam epitaxy.}
\pacs{05.70.Ln}{Nonequilibrium and irreversible thermodynamics.}

\begin{document}

\maketitle

\begin{abstract} 
Conserved growth models that exhibit a nonlinear instability in which
the height (depth) of isolated pillars (grooves) grows in time are
studied by numerical integration and stochastic simulation. When this
instability is controlled by the introduction of an infinite series of
higher-order nonlinear terms, these models exhibit, as function of
a control parameter, a non-equilibrium phase transition between a
kinetically rough phase with self-affine scaling and a phase that
exhibits mound formation, slope selection and power-law coarsening.
\end{abstract}

The nonequilibrium kinetics of the growth of films by the deposition of
atoms on a substrate is of considerable experimental and theoretical
interest\cite{b.rev1,b.rev2}. While the process of kinetic
roughening \cite{b.rev1} leading to a self-affine interface profile has
been extensively studied, there has been much recent
interest\cite{b.rev2,b.mound1,b.mound2} in a different mode of surface
growth, involving the formation and coarsening of ``mounds''
(pyramid-like structures). The system is said to exhibit {\it slope
selection} if the typical slope of the sides of the mounds remains
constant during the coarsening process. 
Traditionally, the formation of mounds has been attributed to the
presence of an Ehrlich-Schwoebel (ES) step-edge barrier\cite{b.es1,b.es2}
that hinders the downward motion of atoms across the edge of a step.
The destabilizing effect of the resulting ``uphill'' surface current is
usually modeled in continuum growth equations\cite{b.sp,b.pv} as a {\it
linear instability} arising from a Laplacian of the height variable
with a negative coefficient. 
%If this uphill surface current is
%compensated by higher-order terms in the growth equation at a specific
%value of the slope, then that slope is selected by the system.

In this Letter, we show that mound formation and power-law coarsening
with slope selection occurs in a class of well-known,
conserved surface growth models as a result of a {\it nonlinear instability}
which leads to a {\it dynamical phase transition} between kinetically
rough and mounded morphologies. We consider the conserved,
fourth-order, nonlinear growth equation proposed by Lai and
Das Sarma\cite{b.lds} and by Villain\cite{b.villain}:
\begin{equation}
\partial h^{\prime}({\bf r},t^\prime)/\partial t^\prime = -\nu \nabla^4
h^{\prime} + \lambda^\prime \nabla^2 |{\bf \nabla}h^{\prime}|^2 + 
\eta^\prime({\bf r},t^\prime), \label{e.lds1}
\end{equation}
where $h^{\prime}({\bf r},t^\prime)$ represents the height variable at the
point ${\bf r}$ at time $t^\prime$, ${\bf \nabla}$ and $\nabla ^2$
represent, respectively, the spatial derivative and Laplacian
operators in $d$ dimensions (the dimension of the substrate),
and $\eta^\prime$ is a Gaussian, delta-correlated random noise.
This equation is believed\cite{b.rev2,b.lds,b.villain} to provide a correct
description of the scaling behavior of kinetically rough surfaces of
films grown by molecular beam epitaxy. Our results are based on
numerical integration of this 
equation in one dimension, using a simple Euler scheme\cite{b.us1}. 
Choosing appropriate units of height, length and time, and discretizing
both space and time, Eq.(\ref{e.lds1}) is written as\cite{b.us1}
\begin{eqnarray}
h_i(t + \Delta t) - h_i(t) &=& \Delta
t \tilde{\nabla}^{2} [ - \tilde{\nabla}^{2} h_i(
t) + 
\lambda |\tilde{\nabla} h_i(t)|^2] \nonumber \\
&+&\sqrt{\Delta t}\, \eta_i(t), \label{e.lds2}
\end{eqnarray}
where $h_i(t)$ represents the
dimensionless height variable at the lattice point $i$ at
dimensionless time $t$, $\tilde{\nabla}$ and
$\tilde{\nabla}^{2}$ are lattice versions\cite{b.us1} of the derivative
and Laplacian operators, and $\eta_i(t)$ is a random
variable with zero average and variance equal to unity. 
These equations, with an appropriate choice of $\Delta t$, 
are used to numerically follow the time evolution of the interface.
We have also studied an atomistic model\cite{b.kds}
in which the height variables $\{ h_i \}$ are integers. 
This model is defined by the following deposition
rule. First, a site (say $i$) is chosen at random. Then the
quantity 
\begin{equation}
K_i(\{ h_j \}) =
-\tilde{\nabla}^{2}h_i + \lambda |\tilde{\nabla}h_i|^2
\label{e.kds1} 
\end{equation}
is calculated for the site $i$ and all its nearest neighbors. Then, a
particle is added to the site that has the smallest value of $K$ among
the site $i$ and its nearest neighbors. In the case of a tie for the
smallest value, the site $i$ is chosen if it is involved in the tie;
otherwise, one of the sites involved in the tie is chosen randomly.
The number of deposited layers provides a measure of time in this model.

It was found earlier\cite{b.us1} that both these models exhibit a {\it
nonlinear instability} in which isolated structures (pillars for
$\lambda >0$, grooves for $\lambda<0$) grow rapidly if
their height (depth) exceeds a critical value. This instability can be
controlled\cite{b.us1} by replacing $|\tilde{\nabla} h_i|^2$ in
Eqns.(\ref{e.lds2}) 
and (\ref{e.kds1}) by $f(|\tilde{\nabla} h_i|^2)$ where the nonlinear
function $f(x) \equiv (1-e^{-cx})/c$, $c>0$ being a control parameter.
We call the resulting models ``model I'' and ``model II'', respectively. 
This replacement, which amounts to the introduction of an infinite series of 
higher-order nonlinear terms, is physically meaningful. 
Politi and Villain\cite{b.pv} have shown that the nonequilibrium
surface current that leads to the $\nabla^2 |{\bf \nabla}h^{\prime}|^2$
term in Eq.(\ref{e.lds1}) should be proportional to
${\bf \nabla}|{\bf \nabla} h^\prime|^2$ when $|{\bf \nabla} h^\prime|$ 
is small, and should go to zero when $|{\bf \nabla} h^\prime|$ is
large. The introduction of the
``control function'' $f(|\tilde{\nabla} h_i|^2)$ satisfies this
physical requirement.

The time evolution of the height variables in model I is,
thus, given by 
\begin{eqnarray}
h_i(t + \Delta t) &-& h_i(t) = \Delta
t \tilde{\nabla}^{2} [ - \tilde{\nabla}^{2} h_i(t) \nonumber \\
&+& \lambda (1-e^{-c|\tilde{\nabla} h_i(t)|^2})/c]
+\sqrt{\Delta t}\, \eta_i(t). \label{e.cld}
\end{eqnarray}
It was shown in Ref.\cite{b.us1} that if $c$ is sufficiently large, then
the instability is absent and both models exhibit the scaling behavior
expected in kinetic roughening. In the present work, we show that as
the value of $c$ is decreased, these models exhibit a {\it first-order
dynamical phase transition}\cite{b.noneq} to a mounded morphology at a
critical value of $c$. The mounded phase exhibits power-law coarsening
(interface width $W \sim t^{\beta^\prime}$), while the slope of the
mounds remains constant. We present results for the phase diagram of
these models in the $(\lambda,c)$ plane and describe the results of a
stability analysis that provides an understanding of the observed
behavior. We also show that this mechanism of mound formation is {\it
qualitatively} different from the conventional ES mechanism.
Our results are quite general in that the behavior found in our
models {\it does not} depend crucially on the form of the function
$f(x)$: any monotonic function that is linear for small $x$ and
saturates for large $x$ leads to similar results. In particular, we have
found very similar behavior using the form,
$f(x)=x/(1+cx)$, suggested by Politi and Villain\cite{b.pv}.

\begin{figure}
\twofigures[width=7cm]{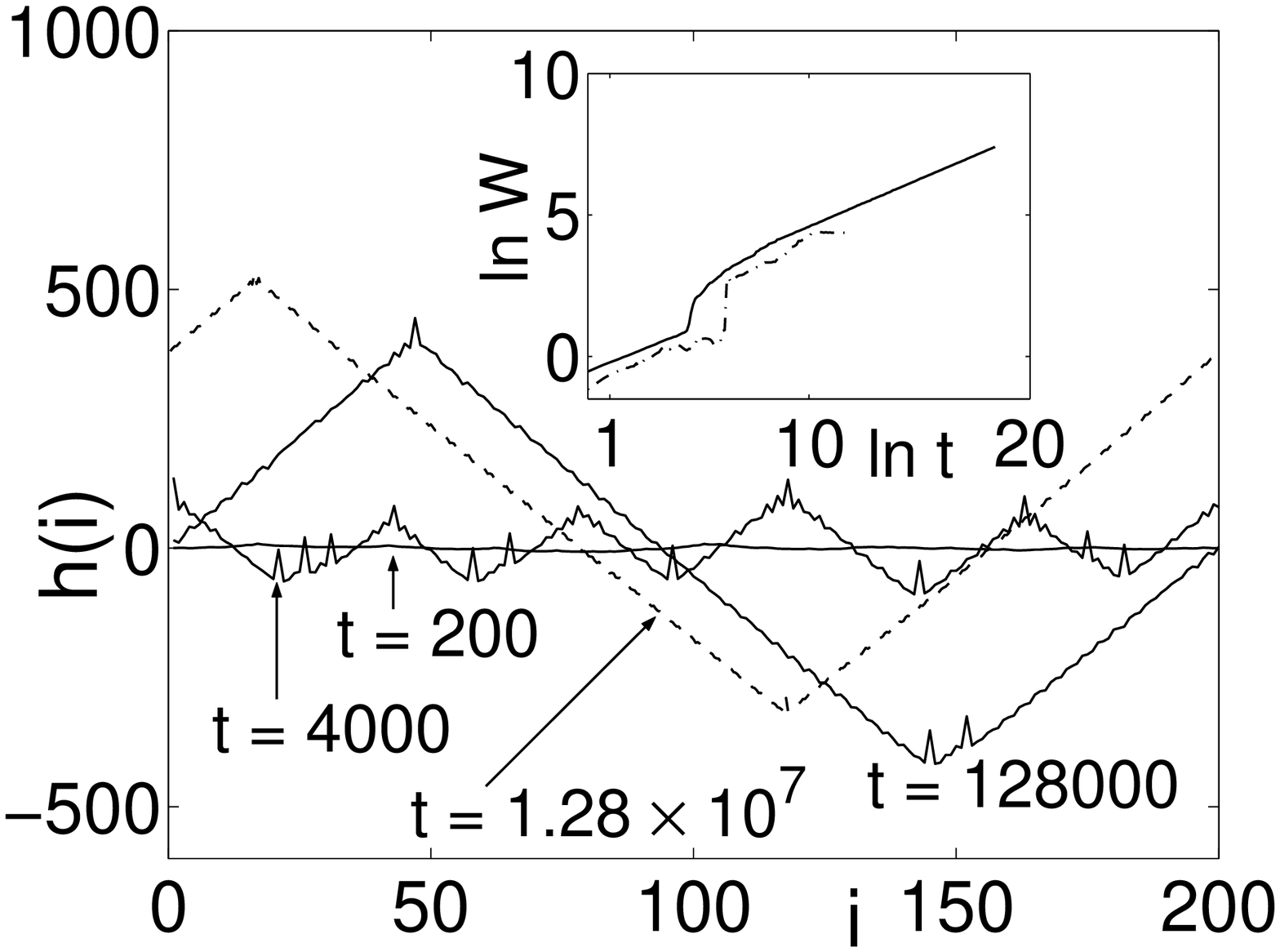}{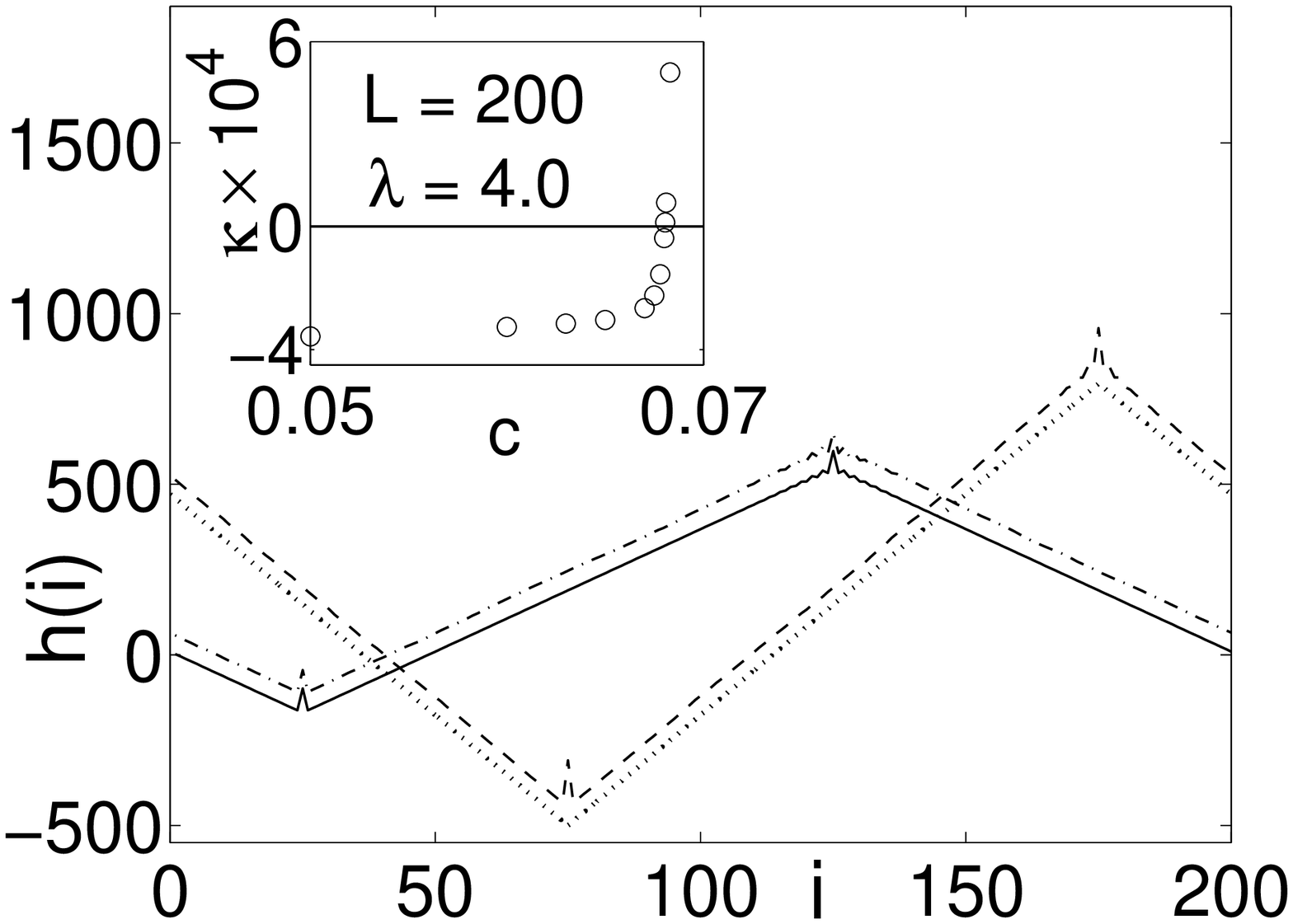}
\caption{Interface profiles at three different times
($t$ = 200, 4000, and 128000) in a run starting from a flat state
for a $L$ = 200 sample of model I with $\lambda=4.0$
and $c=0.02$ (full lines).
A profile for $L = 500$ at $t = 1.28\times 10^7$ for
the same parameters is also shown (dashed line) with
both axes scaled by 2.5.
Double-log plots of the interface width $W$ as a function of time $t$
in the $L=200$ run (dash-dotted line), and similar data averaged over 40 runs
for $L=1000$ samples (full line) are shown in the inset.}
\label{f.fig1}
\caption{Invariant height-profile for
a $L=200$ system with $\lambda=4.0$, $c=0.02$ (full line), and a snap
shot of the same system in steady state (dash-dotted line). 
A snap shot
of a $L=200$ system with $f(x)=x/(1+cx)$, $\lambda=4.0$, $c=0.01$ in the steady state (dashed line) and an invariant profile of the corresponding continuum 
equation (dotted line) are also shown.
Inset: Zero-crossing of the largest eigenvalue $\kappa$ of the stability matrix
as a function of $c$ ($\lambda=4.0$, $L=200$).}
\label{f.fig2}
\end{figure}

Our results are obtained for systems of different sizes ($40 \le L \le
1000$ -- we do not find any significant dependence of the results on
$L$) with periodic boundary conditions. In most of our studies of
model I, we used $\Delta t = 0.01$. We have checked that very similar 
results are obtained for smaller values of $\Delta t$.
If the control parameter $c$ is sufficiently large,
then the nonlinear instability is completely suppressed and the models
exhibits the usual dynamical scaling behavior with the
expected\cite{b.lds} exponent values, $\beta \simeq 1/3$, the dynamical
exponent $z \simeq 3$, and the exponent $\alpha = \beta z \simeq 1$. As
the value of $c$ is decreased with $\lambda$ held constant, the
instability makes its appearance: the height $h_0$ of an isolated
pillar (for $\lambda >0$) increases in time if $h_{min}(\lambda,c) <
h_0 < h_{max}(\lambda,c)$. The value of $h_{min}$ is nearly
independent of $c$, while $h_{max}$ increases as $c$ is decreased. If
$c$ is sufficiently large, $h_{max}$ is small and the instability does
not affect the scaling behavior of global quantities such as $W$.
%although transient multiscaling at length scales shorter than the
%correlation length $\xi \sim t^{1/z}$ may be found\cite{b.us1}. 
As $c$ is decreased further, $h_{max}$ becomes large, and when 
isolated pillars with $h_0 >h_{min}$ are created at
an initially flat interface through random fluctuations, the rapid
growth of such pillars to height $h_{max}$ leads to a sharp upward
departure from the power-law scaling of $W$ with time $t$.
The time at which this departure occurs varies from run to run. Typical
results obtained for model I with $\lambda=4.0$ and $c=0.02$ are shown
in the inset of Fig.\ref{f.fig1}.  

The instability leads to the formation of a large number of pillars of
height close to $h_{max}$. As the system evolves in time, the
interface self-organizes to form triangular mounds of a fixed slope
near these pillars. These mounds then coarsen in time, with large
mounds growing larger at the expense of small ones. In this coarsening
regime, a power-law growth of $W$ in time is recovered.
The slope of the sides of the triangular mounds remains
constant during this process. Finally, the system reaches a steady
state with one peak and one trough and remains in this state for longer
times. 

This behavior is illustrated in Fig.\ref{f.fig1} where the interface
profiles in a typical run for a $L=200$ system starting from a flat
state are shown at times $t=200$ (before the onset of the instability),
$t=4000$ (after the onset of the instability, in the
coarsening regime), and $t=128000$ (in the final steady state).
The inset  shows the time-evolution of $W$ in this run, as well as the 
results for $W$ as a function of $t$, averaged over 40 runs for
$L=1000$ samples. The averaged data show a power-law {\it growth
regime} with $\beta \simeq 1/3$ before the onset of the instability, 
and a second power-law {\it coarsening regime} with $W \sim t^{\beta^\prime}$,
$\beta^\prime = 0.34 \pm 0.01$, at long times. The selection of a
``magic slope'' during the coarsening process is clearly seen in the
plots of Fig.\ref{f.fig1}. More quantitatively, the distribution
of the nearest-neighbor height differences $s_i \equiv |h_{i+1}-h_i|$ is
found to exhibit a pronounced peak at the selected value of the slope,
and the position of this peak does not change during the coarsening
process. The steady-state profile for a $L=500$ sample, also shown in
Fig.\ref{f.fig1}, illustrates the sample-size independence of the results.  
The peak and the trough in the
steady state are separated by $\simeq L/2$ if the boundary condition
requires the heights at the two ends of the sample to be the same.
The occurrence of a peak and a symmetrically placed trough in the
steady state is a consequence of using periodic
boundary conditions. 
This symmetry is not present when other boundary conditions (such as
``fixed'' and ``zero flux'') are used. 
However, the basic phenomenology of
mound formation, power-law coarsening and slope selection does not
depend on the boundary conditions.

While the saw-tooth-like surface profiles found for small $c$ is
qualitatively different from the self-affine morphology observed for
large $c$, the interface width exhibits very similar power-law
behavior in the two cases ($\beta \simeq \beta^\prime \simeq 1/3$, and
$\alpha^\prime = 1 \simeq \alpha$). Thus, a measurement of the
interface width would not distinguish between the two
growth modes. A clear distinction between the two
morphologies may be obtained from measurements of the average number
of extrema of the height profile\cite{b.tor}. The steady-state profile
in the mound-formation regime exhibits two extrema for {\it all} values of the
system size $L$. In contrast, the number of extrema in the steady state
in the kinetic roughening regime increases with $L$ as a power 
law\cite{b.tor} -- we find that for values of $c$ for which the system
is kinetically rough, e.g. for $\lambda = 4.0$, $c=0.05$, 
the average number of extrema in the steady state is proportional to
$L^\delta$ with $\delta \simeq 0.83$. This observation allows us to
define an ``order parameter'' that is zero in the large-$c$, kinetic
roughening regime and finite in the small-$c$, mound-formation regime. Let
$\sigma_i$ be an Ising-like variable, equal to the sign of the slope of
the interface at site $i$. An extremum in the height
profile then corresponds to a ``domain wall'' in the configuration of
the $\{\sigma_i\}$ variables. Since there are two
domain walls separated by $\sim L/2$ in the steady state in the 
mound-formation regime, the quantity
\begin{equation}
m = \frac{1}{L} |\sum_{j=1}^L \sigma_j e^{2\pi i j/L}|, \label{e.op}
\end{equation}
would be finite in the $L \to \infty$ limit. On the other hand, $m$
would go to zero for large $L$ in the kinetically
rough regime because the number of domains in the steady-state profile
would increase with $L$. We find numerically that $m \sim L^{-\gamma}$
with $\gamma \simeq 0.2$ for $\lambda = 4.0$, $c=0.05$.
%%%%%%%%%%%%%%%%%%%%%%%%%%%%%%%%%
%%%%%%%%%%%%%%%%%%%%%%%%%%%%%%%%%%%%%%%%%%%%%%

The behavior described above may be understood from a simple stability
analysis. The profile near the top ($i=i_0$) of a mound may be
approximated as $h_{i_0}=x_0+x_1, \\
h_{i_0 \pm j} = x_0 -|j-1|x_2$, where $x_1$ is the height of the 
pillar at the top of the mound and $x_2$ is the selected slope. 
The condition that this profile does not change under the dynamics 
of Eq.(\ref{e.cld}) with no noise leads to the following pair of 
non-linear equations for $x_1$ and $x_2$:
\begin{eqnarray}
2x_1-\lambda[1-e^{-cx_2^2}]/c &=& 0, \nonumber \\
3x_1 - x_2 - \lambda [1-e^{-c(x_1+x_2)^2/4}]/c &=& 0.
\label{e.stability}
\end{eqnarray}
These equations admit a non-trivial solution for sufficiently small
$c$, and the resulting values of $x_1$ and $x_2$ are found to be quite
close to the results obtained from numerical integration. A similar analysis
for the profile near the bottom of a trough (this amounts to replacing
$x_2$ by $-x_2$ in Eq.(\ref{e.stability})) yields
slightly different values for $x_1$ and $x_2$. The full stable profile
(a fixed point of the dynamics without noise)
with one peak and one trough may be obtained numerically by calculating
the values of $h_i$ for which $g_i = 0$ for all $i$,
where $g_i$ is the term multiplying $\Delta t$ in the right-hand side of
Eq.(\ref{e.cld}). This calculation shows that the small mismatch
between the values of $x_2$ near the top and the bottom is accommodated
by creating a few ripples near the top. The numerically obtained
fixed-point profile for a $L=200$ system with $\lambda=4.0$, $c=0.02$ is
shown in Fig.\ref{f.fig2}, along with a typical steady-state profile for
the same system. The two profiles are found to be nearly identical.
Fig.\ref{f.fig2} also shows a snap shot
of a $L=200$ system with $f(x)=x/(1+cx)$, $\lambda=4.0$, $c=0.01$ in the 
steady state and an invariant profile of the corresponding 
continuum equation, obtained using the procedure of Ref.\cite{b.racz}. These
results show that the steady-state properties for the two forms of $f(x)$
are very similar, and the continuum equation admits invariant solutions
that are very similar to those of the discretized models.

The local stability of the mounded fixed-point may be determined from a
calculation of the eigenvalues of the matrix $M_{ij}=\partial
g_i/\partial h_j$ evaluated at the fixed point. We find that the
largest eigenvalue of this matrix crosses zero at $c=c_1(\lambda)$ (see
inset of Fig.(\ref{f.fig2})), signaling an instability of the mounded
profile. The structure of Eq.(\ref{e.cld}) implies that $c_1(\lambda)
\propto \lambda^2$. Thus, for $0<c<c_1(\lambda)$, the dynamics of
Eq.(\ref{e.cld}) without noise admits two locally stable invariant
profiles: a trivial, flat profile with $h_i$ the same for all $i$, and a
non-trivial one with one mound and one trough. Depending on the initial
state, the no-noise dynamics takes the system to one of these two fixed
points. For example, an initial state with one pillar on a flat
background is driven by the no-noise dynamics to the flat fixed point if the
height of the pillar is smaller than a critical value, and to the
mounded one otherwise. We did not carry out a stability analysis of 
mounded fixed-point solutions of the continuum equation (see Fig.\ref{f.fig2})
because doing such analysis {\it without discretizing space} would be
extremely difficult. 

\begin{figure}
\twofigures[width=7cm]{fig3.eps}{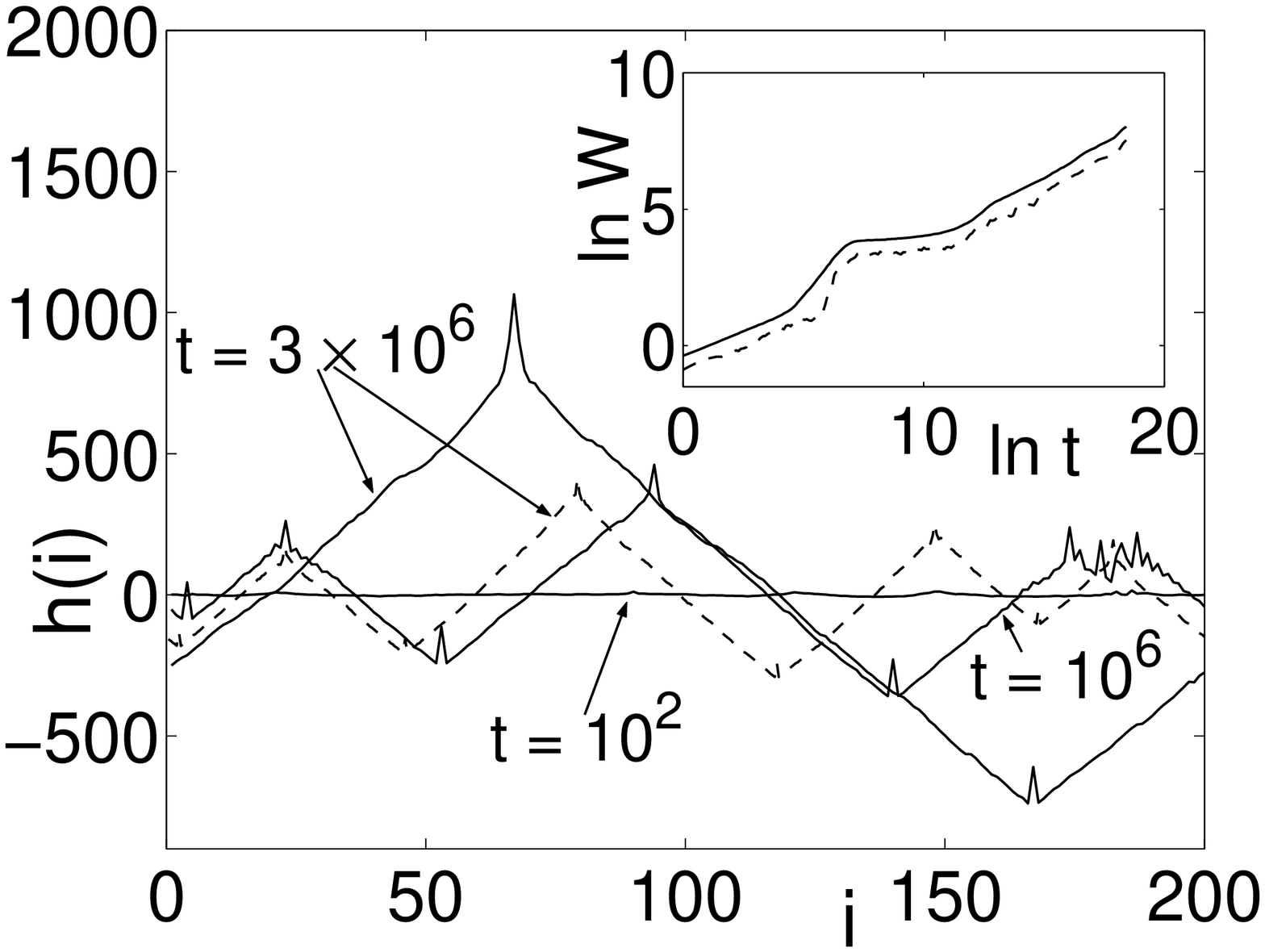}
\caption{Critical values of the control parameter $c$ as functions of 
$\lambda$: $c_1$ of model I (circles), $c_2$ of model I (triangles), 
and $c_2$ of model II (diamonds). Inset: The probabilities $P_1$ (circles) 
and $P_2$ (inverted triangles) defined in text, as functions of $c$ for
model I with $\lambda=4.0$, $L=200$.}
\label{f.fig3}
\caption{Interface profiles at three times ($t = 10^2$,\, $10^6$, and $3.10^6$)
in a run starting from a flat state for
a $L = 200$ sample of model II with $\lambda=2.0$ and $c=0.005$ (full
lines). A profile for $L = 500$ at $t = 3.10^6$ for
the same parameters is also shown (dashed line) with both axes scaled
by 2.5. Double-log plots of the
interface width $W$ as a function of time $t$ in the $L=200$ run (dashed
line), and similar data averaged over 40 runs for $L=1000$ samples (full
line) are shown in the inset.}
\label{f.fig4}
\end{figure}

In the presence of the noise, the perfectly flat fixed point transforms
to the kinetically rough steady state, and the non-trivial fixed point
evolves to the mounded steady state shown in Fig.\ref{f.fig1}. A
dynamical phase transition at $c=c_2(\lambda) < c_1(\lambda)$ separates
these two kinds of steady states. To calculate $c_2(\lambda)$, we start
a system at the mounded fixed point and follow its evolution according to
Eq.(\ref{e.cld}) for a long time (typically $t=10^4$) to check whether it
reaches a kinetically rough steady state. By repeating this
procedure many times, the probability, $P_1(\lambda,c)$, of
a transition to a kinetically rough state is obtained. For fixed $\lambda$,
$P_1$ increases rapidly from 0 to 1 as $c$ is increased above a
critical value. Typical results for $P_1$ as a function of $c$ for
$\lambda=4.0$ are shown in the inset of Fig.\ref{f.fig3}. The value of
$c$ at which $P_1=0.5$ provides an estimate of $c_2$. Another estimate
is obtained from a similar calculation of $P_2(\lambda,c)$, the
probability that a flat initial state evolves to a mounded steady state.
As expected, $P_2$ increases sharply from 0 to 1 as
$c$ is decreased (see inset of Fig.\ref{f.fig3}), and the value of $c$ at
which this probability is 0.5 is slightly lower than the value at which
$P_1=0.5$. This difference reflects finite-time hysteresis effects. The
value of $c_2$ is taken to be the average of these two estimates, and
the difference between the two estimates provides a measure of the
uncertainty in the determination of $c_2$. The phase boundary obtained
this way is shown in Fig.\ref{f.fig3}, along with the results for
$c_1(\lambda)$.
%%%%%%%%%%%%%%%%%%%%%%%%%%%%%%%%%
%%%%%%%%%%%%%%%%%%%%%%%%%%%%%%%%%%%%%

Our numerical results for model II are very similar. Height profiles of
a $L=200$ sample with $\lambda=2.0$, $c=0.005$ at three times (before
the onset of the instability, during coarsening, and at the steady
state) are shown in Fig.\ref{f.fig4}, along with the profile of a $L=500$
sample in the coarsening regime. The inset shows the time-evolution of
the interface width in the $L=200$ run and also the average over 40 runs for
$L=1000$ samples. The phase diagram for this model (see Fig.\ref{f.fig3}) is
very similar to that of model I, while the coarsening exponent appears
to have a higher value, $\beta^\prime = 0.50 \pm 0.01$.

The qualitative behavior found here is very similar to that in
canonical first-order transitions in two- and three-dimensional 
equilibrium systems. In a
time-dependent Ginzburg-Landau description\cite{b.ma} of the dynamics of
a system exhibiting a first-order transition, the no-noise dynamics
exhibits two locally stable fixed points (corresponding to the
uniformly ordered and disordered states) in a range of temperatures
near the transition temperature. In the presence of noise, the system
selects one of the phases corresponding to these two fixed points,
except at the transition temperature where both phases coexist.  The
local stability of the mean-field ordered and disordered states leads
to finite-time hysteresis effects near the transition temperature. The
behavior we find is very similar, with the rough and the mounded states
corresponding to the disordered and the ordered states of equilibrium
systems and $c$ playing the role of the temperature. In analogy with
the behavior of equilibrium systems, we find hysteresis and coexistence
of rough and mounded morphologies near $c=c_2$. 

In summary, we have shown that a nonlinear instability in a 
class of surface growth models leads to mound formation and
power-law coarsening with slope selection via a dynamical phase
transition. This mechanism of mound 
formation is very different from the conventional ES mechanism. The ES
instability is usually modeled\cite{b.sp,b.pv} as a {\it linear} one
arising from a $\nabla^2h$ term with a negative coefficient. Such a
term is clearly absent in our models. Consequently, a flat interface is
{\it locally stable} in our models. Since the non-equilibrium surface current
in our models vanishes for {\it all} values of a constant slope, the
slope selection we find is a true example of nonlinear pattern
formation. In contrast, slope selection occurs in ES-type models only
when the surface current vanishes at a {\it specific value} of 
the slope\cite{b.sp}. 

In view of the observation~\cite{b.us1,b.nb} that spatial 
discretization may drastically affect
the behavior of continuum growth equations, it is interesting to enquire
whether the truly continuum Eq.(\ref{e.lds1}) with $|\nabla h^\prime|^2$ 
replaced by $f(|\nabla h^\prime|^2)$ would exhibit the same behavior as 
that found for the discrete models studied here. While we can not provide a 
rigorous answer to this question, we expect the continuum equation to behave 
in a similar way. It was shown in Ref.\cite{b.us1} that the nonlinear 
instability of the discretized Eq.(\ref{e.lds2}) 
can not be avoided by making the integration time step
sufficiently small, or by using more accurate left-right symmetric definitions of
the lattice derivatives. There is no {\it a priori} reason to believe that
certain {\it left-right asymmetric} discretization schemes~\cite{b.putk} that 
avoid this instability provide a more accurate 
representation of the behavior of
the continuum equation. Since the instability does not necessarily imply a
divergence of the height~\cite{b.us1}, the rigorous proof~\cite{b.putk} 
of the boundedness  of the solutions of Eq.(\ref{e.lds1}) without 
noise does not rule out its occurrence
in the continuum equation. Finally, the existence of mounded fixed-point
solutions of the continuum equation (see Fig.\ref{f.fig2}) 
strongly suggests that its behavior is not
qualitatively different from that of the discrete models.

%%%%%%%%%%%%%%%%%%%%%
\acknowledgments
We thank S. Das Sarma for helpful discussions and SERC, IISc for
computational facilities.
%%%%%%%%%%%%%%%%%%%%%%%

\end{document}